# Distance function wavelets – Part II: Extended results and conjectures

W. Chen

Simula Research Laboratory, P. O. Box. 134, 1325 Lysaker, Norway

E-mail: wenc@simula.no

(21 May 2002)

**Summary**

Report II is concerned with the extended results of distance function wavelets (DFW). The fractional DFW transforms are first addressed relating to the fractal geometry and fractional derivative, and then, the discrete Helmholtz-Fourier transform is briefly presented. The Green second identity may be an alternative devise in developing the theoretical framework of the DFW transform and series. The kernel solutions of the Winkler plate equation and the Burger's equation are used to create the DFW transforms and series. Most interestingly, it is found that the translation invariant monomial solutions of the high-order Laplace equations can be used to make very simple polynomial DFW series. In most cases of this study, solid mathematical analysis is missing and results are obtained intuitively in the conjecture status.

## 1. Introduction

This report is the second in series [1,2] about my latest advances on the distance function wavelets (DFW). Unlike the common distance functions, e.g., MQ and TPS, which have no provision for scaling and carry out a multiresolution hierarchy by simply dropping or adding some points [3], the DFW is comprised of both the scale and translation arguments. To better understand what will be presented here, the readers are advised to have a look at Report I [1] beforehand. The report is featured with lots of grand conjectures, where firm mathematical underpinnings are conspicuously lacking in most cases. But nevertheless the author assumes that many results have certain physical grounds and are in agreement with the faith that God rules the world with simplicity and beauty.

The rest of report is organized into six thematic sections. In section 2, the fractional Helmholtz-Fourier transform (HFT) and series and Helmholtz-Laplace transform (HLT) are presented in relation to fractal geometries and fractional derivatives. In section 3, we briefly present the discrete Helmholtz-Fourier transform without mathematical justifications. Section 4 tries to derive the distance function wavelets by the Green second identity and the Laplace transform. In section 5, the solutions of the Winkler plate equation and the Burger's equation are utilized to create some novel DFW transforms and series. In section 6, the translation invariant monomial solutions of the high-order Laplace equations are applied to develop the polynomial DFW series. Finally, section 7 provides a few supplementary results on the DFW and points out some potential uses.

## 2. Fractional Helmholtz transforms and series

In recent years, much attention has been attracted to the so-called fractional derivative [4], fractional Fourier [5] and Laplace transforms and fractal geometry. The underlying relationships between them are also unraveled [5-7]. Note that "fractional" is just conventional misnomer since it also actually indicates real number relating to the so-called fractional derivative and integral transforms. The HFT and HLT [1] were recognized as the distance function counterparts of the Fourier and Laplace transforms, respectively. By analogy with the latter two, this section addresses the fractional DFW Helmholtz transforms and series.

The HFT is given by

$$F_n(\omega,\xi) = \frac{1}{\sqrt{C_n}} \int_{IR^n} f(x) E_n^{-i\omega\|\xi-x\|} dx \tag{1a}$$

and

$$f(x) = \frac{1}{\sqrt{C_n}} \int_{-\infty}^{+\infty} \int_{IR^n} F_n(\omega,\xi) E_n^{i\omega\|x-\xi\|} d\xi d\omega, \tag{1b}$$

while the HLT [1] is stated by

$$L_n(s,\xi) = \int_{IR^n} f(x) E_n^{-s\|\xi-x\|} dx, \tag{2}$$

where $n$ denotes the dimensionality, and the function $E$ [1] is defined by

$$E_n^{-\lambda x} = \begin{cases} \dfrac{\lambda^{1/2}}{2\pi} e^{-\lambda x}, & n = 1, \\ \dfrac{\lambda^{n-1/2}}{2\pi} (2\pi\lambda x)^{-(n/2)+1} K_{(n/2)-1}(\lambda x), & n \geq 2, \end{cases} \quad x \geq 0, \tag{3a}$$

$$E_n^{i\lambda x} = \begin{cases} \dfrac{\lambda^{1/2}}{2\pi} (\cos(\lambda x) + i\sin(\lambda x)), & n = 1, \\ \dfrac{i\lambda^{n-1/2}}{4} (2\pi\lambda x)^{-(n/2)+1} \left(J_{(n/2)-1}(\lambda x) + i Y_{(n/2)-1}(\lambda x)\right), & n \geq 2, \end{cases} \quad x \geq 0, \tag{3b}$$

where $J_{n/2}$ and $Y_{n/2}$ are respectively the Bessel functions of the first and second order of the n/2-1 order, and $K_{n/2}$ is the modified Bessel function of the second kind. There is a number of ways to connect the HFT and HLT with the fractional derivative. One simple way is through frequency domain. Obviously, the HFT hold

$$F[\nabla^2 f(x)] = (i\omega)^2 F[f(x)] = (i\omega)^2 F(\omega,\xi), \tag{4}$$

and generally, this process may be iterated for the *m*-th order divergence derivative to

$$F[\nabla^m f(x)] = (i\omega)^m F[f(x)], \tag{5}$$

where *m* is a integer number in the most stereotyped mathematics. Observing (5), there is no reason hindering the extension of *m* to the set of real numbers, and then, comes out the fractional derivative. If *m*<0 in (5), then we have fractional integration. Let us advance a little further. *m* could be even a complex number corresponding to a complex order partial differentiation [8], while keeping (5). We could develop a complete theory of fractional HFT and HLT by analogy with the fractional Fourier and Laplace transforms.

On the other hand, Blu and Unser [3] have pointed out that the self-similarity of fractal geometry is one fundamental concept to create the distance function wavelet. This inspired the author to assume that dimensionality *n* can be a real number. Consequently, (1) and (2) turn out to be a new type of the fractional HFT and HLT again

$$E_n^{-\lambda x} = \begin{cases} \dfrac{\lambda^{1/2}}{2\pi} e^{-\lambda x}, & n=1, \\ \dfrac{\lambda^{n-1/2}}{2\pi} (2\pi\lambda x)^{-(n/2)+1} K_{(n/2)-1}(\lambda x), & n \neq 1, \end{cases} \quad x \geq 0, \tag{6a}$$

$$E_n^{i\lambda x} = \begin{cases} \dfrac{\lambda^{1/2}}{2\pi}(\cos(\lambda x) + i\sin(\lambda x)), & n = 1, \\ \dfrac{i\lambda^{n-1/2}}{4}(-2\pi i\lambda x)^{-(n/2)+1}\left(J_{(n/2)-1}(\lambda x) + iY_{(n/2)-1}(\lambda x)\right), & n \neq 1, \end{cases} \quad x \geq 0. \qquad (6b)$$

The same is for fractional Helmholtz series given in [1]. A fractional *n* implies fractal geometry. It is noted that the DFW HFT has a clear edge over the classic Fourier transform in that the dimensionality explicitly appears in the HFT and thus may be suitable to serve as an alternative powerful mathematical tool to quantitatively describe the fractal geometry. In some areas (e.g. control engineering), the complex dimension is a useful concept. Thus, *n* even could be a complex number. One thing that deserves more attention is how to define the Euclidean distance in complex dimension geometry, and then how about a negative *n*. The problem is what physical backdrops are behind these exotic mathematical devices. The author conceives that there is an underlying link between the complex dimensionality and complex order derivative through the HFT or the HLT, when *n* is a complex number in (1), (2) and (3). It is certain that there are lots of issues unanswered out there.

The above research displays that the dimensionality (fractal), scale (frequency) and differentiation can be connected through the DFWs. [2] will give a detailed discussion on these inherent relationships through the research of the frequency dependent attenuation of acoustics and elastic waves.

### 3. Discrete Helmholtz-Fourier transform

In terms of continuous Helmholtz-Fourier transform (1), it is easy to attain the 1D discrete Helmholtz-Fourier transform (DHT) as in the discrete Fourier transform. In multidimensional cases, the basis functions of the continuous complex HFT, however, encounter the singularity at the origin. It is noted that the basis functions of the continuous HF J transform

$$\varphi_1(\lambda r_k) = \frac{\lambda^{1/2}}{2}\cos(\lambda r_k), \tag{7a}$$

$$\varphi_n(\lambda r_k) = \frac{\lambda^{n-1/2}}{2\pi}(2\pi\lambda r_k)^{-(n/2)+1} J_{(n/2)-1}(\lambda r_k), \quad n\geq 2 \tag{7b}$$

has arbitrary degree of continuity, where $r_k = \|x - x_k\|$. The HF J transforms are stated as

$$F_n(\lambda, x_k) = \frac{1}{C_J}\int_\Omega f(x)\varphi_n(\lambda\|x_k - x\|)dx \tag{8a}$$

and

$$f(x) = \int_0^{+\infty}\int_\Omega F_n(\lambda, x_k)\varphi_n(\lambda\|x - x_k\|)dx_k d\lambda. \tag{8b}$$

The corresponding discrete transforms are

$$\hat{F}_n(\lambda_j, \xi_l) = \sum_{k=1}^N f(x_k)\varphi_n(\lambda_j\|\xi_l - x_k\|), \tag{9a}$$

and

$$f(x_k) = \frac{1}{N_g}\sum_{j=1}^M \sum_{l=1}^N \hat{F}_n(\lambda_j, \xi_l)\varphi_n(\lambda_j\|x_k - \xi_l\|). \tag{9b}$$

We do not give any mathematical justification in this section.

## 4. Distance function wavelets generated by Green's function

[9] derives various coordinate variable transforms from the Green's function. This section tries to follow the same strategy to develop the distance function transforms. Consider the partial differential equation

$$i\frac{\partial \phi(x,t)}{\partial t} = \Re\{\phi(x,t)\} \qquad (10)$$

with the initial condition

$$\phi(x,0) = f(x) \qquad (11)$$

and boundary conditions

$$\begin{cases} u(x,t) = 0, & x \in S_u, \\ \dfrac{\partial u(x,t)}{\partial n} = 0, & x \in S_T, \end{cases} \quad t \geq 0, \qquad (12)$$

where $\Re$ is the spatial partial differential operator, and $x$ is a multidimensional variable. In terms of the Laplace transform with respect to time $t$, we have

$$[\Re - is]\phi(x,s) = -if(x). \qquad (13)$$

The Green function solution of (13) is

$$\phi(x,s) = -i\int_{\Omega_n} g(x-\xi;-is)f(\xi)d\xi, \qquad (14)$$

where $g(x-\xi,-is)$ is the Green function of the operator $\Re$-$is$ satisfying the boundary condition (12). Applying the inverse Laplace transform to (14) and let $t$=0, we attain

$$f(x) = -\frac{1}{2\pi} \int_{\gamma-i\infty}^{\gamma+i\infty} ds \int_{\Omega_n} g(x-\xi;-is) f(\xi) d\xi. \tag{15}$$

Let $s=i\lambda$, where $\lambda$ are the eigenvalues of the operator $\Re+\lambda$, we have

$$f(x) = \frac{1}{2\pi i} \int_{-ic-\infty}^{-ic+\infty} d\lambda \int_{\Omega_n} g(x-\xi;\lambda) f(\xi) d\xi, \tag{16}$$

If $\Re$ is a self-adjoint operator, all eigenvalues $\lambda$ are real. (16) is restated as

$$f(x) = \frac{1}{2\pi i} \int_{-\infty}^{+\infty} d\lambda \int_{\Omega_n} g(x-\xi;\lambda) f(\xi) d\xi. \tag{17}$$

It is known that the Green functions of many PDEs can be expressed by the corresponding eigenfunctions, i.e.

$$g(x-\xi;\lambda) = \sum_{m=0}^{\infty} \alpha \psi_m(x) \psi_m(\xi), \tag{18}$$

where $\psi_m$ is the eigenfunctions, and the index $m$ is related with different eigenvalues $\lambda$. To proceed further, we need to postulate that due to the translation or rotational invariant, (18) may be restated as

$$g(x-\xi,\lambda) = \frac{1}{C} \int_{\Omega_n} \lambda^{2n-1} \overline{\psi(x-\xi,\lambda)} \psi(\xi-x,\lambda) dx, \tag{19}$$

where the upper bar denotes the complex conjugate. (20) is a DFW expression of $f(x)$. Now substituting (19) into (17) produces the DFW transform

$$f(x) = \frac{1}{C_\psi} \int_{-\infty}^{+\infty} d\lambda \int_{\Omega_n} \psi(\xi-x,\lambda) d\xi \int_{\Omega_n} \lambda^{2n-1} \overline{\psi(x-\xi,\lambda)} f(\xi) dx. \tag{20}$$

On the other hand, it is observed that (17) can also be seen as an anomalous wavelet representation. Namely,

$$F(x,\lambda) = \int_{\Omega_n} g(x-\xi;\lambda) f(\xi) d\xi, \qquad (21a)$$

and

$$f(x) = \frac{1}{2\pi i} \int_{-\infty}^{+\infty} F(x,\lambda) d\lambda. \qquad (21b)$$

For instance, in terms of the free Green function of the Helmholtz equation in 1D infinite domain, the Fourier transform can be restated

$$F(x,\lambda) = \int_{-\infty}^{+\infty} f(\xi) e^{i\lambda(x-\xi)} d\xi, \qquad (22a)$$

and

$$f(x) = \frac{1}{2\pi i} \int_{-\infty}^{+\infty} F(x,\lambda) d\lambda. \qquad (22b)$$

The wavelets transforms (21) and (22) have exotic expressions compared with the standard wavelet formalism. But nevertheless they have most properties of the wavelets.

### 5. Some distance function wavelets

Except the distance function wavelets developed [1], there are plenty of the solutions of various PDEs eligible to create the DFW. This section is dedicated to developing a few new types of the DFW.

**5.1. DFWs involving Kelvin functions**

The deflection of thin elastic plates resting on a Winkler elastic foundation with stiffness $\kappa$ is governed by

$$\nabla^4 u + \kappa^2 u = \begin{cases} -\Delta_i \\ 0 \end{cases} \qquad (23)$$

where $\Delta_i$ represents the Dirac delta function at a source point $i$ corresponding to the fundamental solution (vs. zero for general solution); domain $\Omega$ can be unbounded or bounded with or without boundary conditions. The plate is a two dimensions problem. However, in this study, we extend equation (23) up to five dimensions since we found their fundamental solution [10]

$$\varphi_n^*(r_k) = \frac{i}{2\pi}\left(r_k\sqrt{\kappa}\right)^{-n/2+1}\left(\ker_{n/2-1}(r_k\sqrt{\kappa}) + i\,\mathrm{kei}_{n/2-1}(r_k\sqrt{\kappa})\right), \quad 2 \le n \le 5, \qquad (24)$$

and general solution

$$\varphi_n^\#(r_k) = \left(r_k\sqrt{\kappa}\right)^{-n/2+1}\left(\mathrm{ber}_{n/2-1}(r_k\sqrt{\kappa}) + i\,\mathrm{bei}_{n/2-1}(r_k\sqrt{\kappa})\right), \qquad 2 \le n \le 5, \qquad (25)$$

where ker and kei are respectively the Kelvin and modified Kelvin functions of the second kind, and ber and bei respectively represent the Kelvin and modified Kelvin functions of the first kind. Note that ber and bei have arbitrarily degree of differential continuity, kei has the second differential continuity, but ker encounters a singularity at the origin.

Steady Schrodinger's equation in a radial domain

$$x^2 w'' + x w' - (ix^2 + v^2) w = 0 \qquad (26)$$

also has the solutions (24) and (25). By using the fundamental solution (24), we can construct the distance function transform for a suitable function $f(x)$

$$F(\kappa,\xi) = \int_{IR^n} f(\eta)\overline{\varphi_n^*(\|\xi-\eta\|)}\kappa^{n-1}d\eta, \quad 2\leq n \leq 5, \tag{27a}$$

and

$$f(x) = C_\varphi^{-1}\int_0^{+\infty}\int_{IR^n} F(\kappa,\xi)\varphi_n^*(\|x-\xi\|)d\xi d\kappa, \quad 2\leq n \leq 5, \tag{27b}$$

The solution (25) is also capable to construct the DFW. By analogy with the Helmholtz-Fourier series [1], we have expansion series

$$f(x) = f_0(x) + \sum_{j=1}^{\infty}\sum_{k=1}^{\infty}\left(r_k\sqrt{\kappa_j}\right)^{-n/2+1}\left(\alpha_{jk}ber_{n/2-1}(r_k\sqrt{\kappa_j}) + \beta_{jk}bei_{n/2-1}(r_k\sqrt{\kappa_j})\right),$$

$$2\leq n \leq 5, \tag{28}$$

where $\alpha_{jk}$ and $\beta_{jk}$ are the expansion coefficients. Since $f_0(x)$ is related to zero value of $\kappa$, equation (23) degenerates into a biharmonic equation

$$\nabla^4 f_0(x) = 0 \tag{29}$$

with the boundary data. As we did for the Helmholtz-Fourier series [1], $f_0(x)$ can be evaluated by the boundary element method or the boundary knot method.

**5.2. DFW transform and series with solutions of Burger's equation**

The Berger's equation [11] for large deflections of plate is

$$\nabla^4 u - \alpha^2 \nabla^2 u = \begin{cases} -\Delta_i \\ 0 \end{cases} \tag{30}$$

where the Berger parameter $\alpha$ is constant over the domain but nonlinearly depends on the lateral load. The nonlinear relation between the external load and the deformation is represented by the second right-hand term [11]. We have the fundamental solution [12]

$$\psi_2^*(r_k) = \frac{-1}{2\pi\alpha^2}\left(\ln(r_k) + K_0(\alpha r_k)\right), \tag{31a}$$

$$\psi_n^*(r_k) = \frac{r_k^{2-n}}{(n-2)S_n(1)} + \left(\frac{\alpha}{2\pi r_k}\right)^{-n/2+1} K_{n/2-1}(\alpha r_k), \qquad 3 \leq n, \tag{31b}$$

and general solution [11]

$$\psi_n^\#(r_k) = \left(1 + \left(\frac{\alpha}{2\pi r_k}\right)^{-n/2+1} I_{n/2-1}(\alpha r_k)\right), \qquad 2 \leq n, \tag{32}$$

where $K_{n/2-1}$ and $I_{n/2-1}$ are respectively the modified Bessel functions of the first and second kinds, and $S_n(1)$ is the surface size of unit n-dimensional sphere. By analogy with the Helmholtz-Laplace transform, we have

$$B(\alpha, \xi) = \int_{IR^n} f(\eta)\psi_n^*(\alpha\|\xi - \eta\|)\alpha^{2n-1} d\eta. \tag{33a}$$

$$f(x) = \frac{1}{C_\psi} \int_{\gamma-i\infty}^{\gamma+i\infty} \int_{IR^n} B(\alpha, \xi)\psi_n^\#(\mu\|x - \xi\|) d\xi d\alpha. \tag{33b}$$

The general solution (32) can also be used to construct the DFW series for representing the functions bounded within finite domains.

## 6. DFW series with monomial Laplacian solutions

There are numerous solutions of the high-order Laplace equations satisfying rotational or translation invariant. This section will develop the distance function wavelets series by using the translation invariant monomial solutions. In [2], the rotational invariant solutions are employed to create the DFW.

A translation in the plane is a transform

$$x_i = x + a, \qquad y_i = y + b. \tag{34}$$

Translation invariant with the Laplacian [13] means

$$\nabla^2 (x_i + y_i) = \nabla^2 (x + y) . \tag{35}$$

For *p*-order Laplacian, invariance holds for *m*-order translation when *m*<2*p*, i.e. under

$$x_i^m = (x+a)^m, \qquad x_i^m y_i^m = (x+a)^m (y+b)^m, \qquad y_i^m = (y+b)^m, \tag{36}$$

we have

$$\nabla^{2p} (x_i^m + x_i^m y_i^m + y_i^m) = \nabla^{2p} (x^m + x^m y^m + y^m) . \tag{37}$$

Any real function satisfying the Laplacian is called a harmonic function. In this study, we call the function, which satisfies the high-order Laplacian, the high-order harmonic function. The harmonic function is also called the potential function which includes the scalar and vector potential functions in engineering. Thus, the high-order harmonic function is also often called the high-order scalar or vector potential functions. A linear combination of translates of the monomial high-order harmonic functions can approximate many smooth functions under the harmonic basis function space. As an illustrative example, let us consider the two-dimension translation-invariant monomial solutions of high order Laplace equations

$$\{1, x - x_k, y - y_k, (x - x_k)^2, (x - x_k)(y - y_k), (y - y_k)^2 \ldots\}, \tag{38}$$

we can construct the DFW series to approximate continuously differential function $Q(x,y)$

$$Q(x, y) = c_0 + \sum_{k=1}^{\infty} \left[ c_{1k}(x - x_k) + c_{2k}(y - y_k) + c_{3k}(x - x_k)^2 \right. \\ \left. + c_{4k}(x - x_k)(y - y_k) + c_{5k}(y - y_k)^2 \ldots \right] \tag{39}$$

where $c_{jk}$ are the expansion coefficients. Note that since the high order Laplacian is scale invariant operator, the power exponents of monomials are considered the "scale parameter" here. For higher dimension problems, the similar DFW expansion series can be constructed. (39) is absolutely not a Taylor expansion although it may look like the latter. (39) can be restated as

$$Q(x, y) \cong \sum_{i=0}^{N_x} \sum_{j=0}^{N_y} \sum_{k=1}^{M} c_{ij,k}(x - x_k)^i (y - y_k)^j. \tag{40}$$

Note (40) in fact have only one expansion coefficient for the constant term, i.e. $c_0 = \sum_{k=1}^{M} c_{oo,k}$. In terms of the finite difference, least square, collocation or Galerkin schemes, the polynomial DFW expansion (40) can be simply used to the function approximation and the numerical integration and the solution of partial differential equation under arbitrary domain geometry in an explicit multiscale and meshfree fashion [14]. As in the wavelets, we can truncate the scales and translates locally and get a sparse DFW interpolation matrix in approximating $Q(x,y)$. The completeness, accuracy and numerical tests of the presented polynomial DFW are under study and will be reported in a subsequent paper.

The Laplace equation is typically an elliptical equation. The present polynomial DFW series is thus expected to perform well for a variety of elliptical equations. When applied to parabolic and hyperbolic equations, it would be more efficient and reliable to modify this DFW series (40) via simple function transform to reflect the features of those equations. For example, we can transfer the convection-diffusion equation

$$D\nabla^2 u + \vec{v} \bullet \nabla u - ku = b(x, y, t) \tag{41}$$

into the elliptic modified Helmholtz equation

$$\nabla^2 w - \tau^2 w = 0 \tag{42}$$

by an exponential variable transformation [15,16]

$$u(x) = \exp\left[\frac{-\vec{v} \bullet \vec{r}_k}{2D}\right] w(x), \tag{43}$$

where $\vec{r}_k$ is the distance vector between the source and field points, and

$$\tau = \left[(|\vec{v}|/2D)^2 + \kappa/D\right]^{\frac{1}{2}}. \tag{44}$$

The nonsingular general solution of the convection-diffusion equation (41) is given by

$$u_n^\#(\tau, \vec{r}_k) = \frac{\tau^{n-1/2}}{2\pi} e^{-\frac{\vec{v}\cdot\vec{r}_k}{2D}} (2\pi\tau r_k)^{-(n/2)+1} I_{(n/2)-1}(\tau r_k), \qquad n \geq 2, \tag{45}$$

where *n* is the dimensionality and $r_k$ is the Euclidean distance as defined with (7b). Accordingly, we construct the following modified polynomial DFW expansions

$$u(x,y,t) \cong \sum_{i=0}^{N_x}\sum_{j=0}^{N_y}\sum_{k=1}^{M} a_{ij,k}(t)\exp\left[\frac{-v_x(x-x_k)-v_y(y-y_k)}{2D}\right](x-x_k)^i(y-y_k)^j \quad (46a)$$

and

$$u(x,y,t) \cong \sum_{i=0}^{N_x}\sum_{j=0}^{N_y}\sum_{k=1}^{M} d_{ij,k}(t) u_n^{\#}(\tau,\vec{r}_k)(x-x_k)^i(y-y_k)^j. \quad (46b)$$

The approximation series (46) embeds the velocity direction.

We also can simplify (40) in some way. For example, the normal multiquadratic radial basis function approximation

$$Q(x,y) \cong \sum_{k=1}^{M} h_k \sqrt{(x-x_k)^2+(y-y_k)^2+s_k^2} \quad (47)$$

can be understood a simplified version of (40) with a single scale parameter (shape parameter $s_k$). Determining $s_k$, however, is often tricky and problem dependent.

In addition, the Laplacian DFW series can be constructed under the polar coordinates, i.e.

$$Q(x,y) \cong B_0 + \sum_{j=1}^{N}\sum_{k=1}^{M} r_k^j \left[A_{jk}\sin(j(\theta-\theta_k))+B_{jk}\cos(j(\theta-\theta_k))\right], \quad (48)$$

where $\theta-\theta_k$ is the angle between vectors $(x,y)$ and $(x_k,y_k)$. $j$ indicates the scale. (48) is not easy to use in general because of the difficulty in the evaluation of the angle.

It is feasible to create similar translation invariant DFW series via the translation invariant sine and cosine solutions of the multidimensional Helmholtz equations in handling periodic problems. In 2D case, we have

$$P(x,y) \cong \sum_{i=0}^{N_x} \sum_{j=0}^{N_y} \sum_{k=1}^{M} c_{ij,k} \sin(2\pi i(x-x_k)) \cos(2\pi j(y-y_k)). \tag{49}$$

For the diffusion and convection-diffusion problems, the similar DFW expansion can be made via exp(-$\tau$(x-$x_k$)) and exp(-$\tau$(y-$y_k$)) and their combination with the velocity term as shown in (43). In some particular application, we can construct a special translation invariant distance functions which reflects the systematic characteristics (PDE and outer forcing terms).

It is very desirable to have the orthogonal DFW expansion of the form:

$$u(x,y,t) \cong \sum_{i=0}^{N_x} \sum_{j=0}^{N_y} \sum_{k=1}^{M} b_{ij,k} \phi_i^x(x-x_k) \phi_j^y(y-y_k), \tag{50}$$

where $\phi_i^x$ and $\phi_i^y$ are orthogonal eigenfunctions. It is very interesting to note that unlike the fundamental solution and general solution DFWs in [1] and the preceding sections, the polynomial and trigonometric DFW series developed in this section could have their basis functions from the PDE solution (the separation of the variables) under the rectangular or hypercube domains instead of the circle or hypersphere domains. As such, (50) can nevertheless be simply used to handle arbitrary domain problems due to its translation invariant property. For plenty of PDEs, orthogonal eigenfunctions under rectangular and cube domains can be found in literature. Thus, it is not a difficult task to construct such translate DFW series to solve a broad variety of problems. However, we do not know if these eigenfunctions are still orthogonal under arbitrary domains.

The essential distinction of the present DFW expansions from the other coordinate variable expansion approaches lies in that we use the translation invariant distance variable solutions of the high order PDE to get the meshfree and multiscale interpolation schemes. It is stressed that as the spherically symmetric solutions are closely related to the fundamental and general solutions, most of the coordinate variable kernel solutions of PDEs under regular domains are translate invariant solutions for arbitrary domain problems.

The above polynomial DFW series in space can be extended to the time-space DFW. It is known that the wave and heat problems have the time-harmonic solutions, which are in some sense similar to the spatial harmonic functions of the Laplacian. In particular, the wave equation has the time-reversal invariance solution. We may have the time-space polynomial DFW series

$$u(x,y,t) \cong \sum_{s=0}^{N_t}\sum_{i=0}^{N_x}\sum_{j=0}^{N_y}\sum_{k=1}^{M} a_{sij,k}\phi_s^t(t-t_k)\phi_i^x(x-x_k)\phi_j^y(y-y_k), \qquad (51)$$

where $t-t_k \geq 0$ for the diffusion problem and $c^2(t-t_k)^2 - r_k^2 \geq 0$ for wave problems are required to ensure the principle of casualty.

To illustrate the difference between the translation and rotational polynomial DFW series of harmonic functions, we briefly list the rotational invariant Laplacian DFW series [2]

$$Q(x) = Q_0(x) + \sum_{m=1}^{\infty}\sum_{k=1}^{\infty}\alpha_{mk} u_{L_n^m}^*(\|x-x_k\|), \qquad n \geq 2, \qquad (52)$$

where $u_{L_n^m}^*$ is the fundamental solution of the $m$-th order Laplacian $\nabla^{2(m+1)}u$. When the order $m \geq 1$, the high-order Laplacian fundamental solutions of 2D and 3D cases [17] are no longer singular at the origin. $Q_0(x)$ is evaluated via the Green second identity

$$Q_0(x) = \int_{S^{n-1}}\left\{\frac{\partial Q(x_j)}{\partial n}u_{L_n^0}^*(\|x-x_j\|) - Q(x_j)\frac{\partial u_{L_n^0}^*(\|x-x_j\|)}{\partial n}\right\}dx_j, \qquad (53)$$

where $S^{n-1}$ are the surface of finite domains. (53) can be easily evaluated by the boundary element method (BEM) with boundary data, and then $Q_0(x)$ at any inner locations can be calculated via (52). The boundary knot method [10] is also an alternative to the BEM for this task. For more details on the rotational invariant DFW series see the section 5 of [2]. (52) has simpler form and fewer expansion coefficients for high dimension problems than the preceding monomial Laplacian DFW series, and $u^*_{L_n^m}$ of different orders $m$ are orthogonal. The drawback is that the calculation of the Laplacian fundamental solution is more costly than that of monomials.

## 7. Supplementary results

The DFW is easily extended to hand the problems with anisotropic parameters via the modified definition of the distance variable [18]. For instance, consider the Helmholtz equation

$$\nabla(\eta\nabla u) + \lambda^2 u = \begin{cases} -\Delta_i, \\ 0, \end{cases} \quad \text{in } \Omega, \tag{54}$$

where $\eta=\{\eta^j\}$ are the anisotropic parameters in different coordinate directions. If all $\eta^j$ are constants, we have the common fundamental and general solutions via the modified Euclidean distance variable

$$\hat{r}_k = \sqrt{\sum_{j=1}^{n} \eta^j (x^j - x_k^j)^2} \ . \tag{55}$$

Then, we can replace the standard Euclidean variable in the HFT and HLT with the above definition of distance variable. These modified HFT and HLT will be more efficient to handle anisotropic problems.

For nonlinear problems, the distance variable may be defined by

$$\hat{r}_k = \sqrt{\sum_{j=1}^{n} \eta^j (u(x))(x^j - x_k^j)^2} \ . \tag{56}$$

Another useful technique handling nonlinear problems is to use the Kirchhoff transform as in the boundary element method. For detailed discussions of different definitions of the Euclidean distance variable see [18]. For data processing, we need to keep in mind the physical mechanism (PDE model) behind the data and consider the PDE-specific DFW technique.

The distance function wavelets have some potential important applications. For example, the Helmholtz-Fourier transform may be a competitive alternative to the Fourier transform in various data processing techniques such as filtering, modulation and correlation.

The DFW transform and series are established on the kernel distance variable solutions of various linear time-space invariant PDE systems. In more general sense, the strategy can be extended to the gauge invariant PDE model (gauge group) such as Yang-Mills equation. Further, the inverse-scattering transform reveals that invariance (the superposition principle) also holds in some important nonlinear PDE models. We are now under way to research the DFW transform and series based on such nonlinear invariance mechanism. [13] puts it "the invariance has powerful consequences which lie at the heart of the physical idea". We add that "distance" (various relative differences) is an underlying fundamental devise depicting the invariant relativity. Such philosophy implies the distance function wavelets, which underlie invariance and relativity, may be a powerful tool to display, analyze and exploit universe symmetry, simplicity and beauty out of mundane irregularity, clutter and difficulty.

It is worth pointing out again that most results in this report are intuitively attained without any rigorous mathematical analysis. The readers need to note their conjecture status. Further research is now under way and the author would like to get any comments and opinions